# Extending the Maier-Saupe theory to cybotactic nematics

S. Droulias, A.G. Vanakaras and D.J. Photinos*

*Department of Materials Science, University of Patras,
Patras 26500, Greece*

* *photinos@upatras.gr*

**Abstract**
A theory of thermotropic nematic liquid crystals in which molecules form internally ordered clusters is presented. The formulation is based on the same mean field approximation and form of the anisotropic potential used in the Maier-Saupe theory. A uniaxial nematic and two macroscopically isotropic phases are predicted. The lower-temperature isotropic phase consists of thermodynamically stable clusters with internal orientational order. The transition from this phase to the nematic phase is characterized by the divergence of cluster size whilst the entropy and the order parameter change continuously.

**1. Introduction**
The Maier-Saupe (M-S) theory [1-3] is a historical milestone in the theoretical description of thermotropic nematics; the structurally simplest, most common and technologically most important of all types of liquid crystals. Considering its simplicity, the M-S theory is amazingly successful in accounting for the basic features of the nematic-isotropic (N-I) phase transition and of the temperature dependence of the anisotropy exhibited by dielectric, diamagnetic and optical properties in the nematic phase. In their pioneering paper [1], entitled *"A simple molecular theory of the nematic liquid-crystalline state"*, Maier and Saupe wrote *"A theory of nematic liquids must above all provide an explanation for the existence of this long-range order, as well as for the order of magnitude of the degree of order, and of its temperature dependence"*. And this is precisely what their "simple molecular theory" does.

The M-S theory, being based on the mean-field approximation, has all the draw-backs associated with the complete neglect of correlations in the orientations and positions of neighboring molecules; and this is reflected on several qualitative and quantitative deviations of the theoretical predictions from the experimentally observed behaviour. Also, due to the very simple form used for the mean-field potential, the theory predicts universal values for the order parameter and for the entropy change at the N-I transition and a universal temperature dependence of the order parameter in terms of a reduced temperature. Although this is not quite in agreement with what is observed experimentally, the deviations are in general not very large for common thermotropic nematics. Improvement of the agreement with experiment has been obtained either by improving on the mean-field approximation [4] or by using more elaborate forms for the mean-field potential [5]. Aside from such improvements, much attention has been devoted [6,7] to understanding why the theory works so well despite (i) the neglect of molecular correlations, (ii) the assumed dominance of long-range anisotropic dispersion interactions among the molecules and (iii) the isotropic averaging of those interactions over the intermolecular positions. One of the reasons for the successfulness of the M-S theory is that, as pointed out by Gelbart [8,9], the analytical form used for the mean-field is quite general and can accommodate short range interactions as well, particularly intermolecular repulsions of the excluded-volume type. It has also been suggested by De Jeu [7] that the successfulness of the M-S theory is due to mutually compensating approximations made in the derivation of the mean-field. Yet another explanation, suggested by Luckhurst and Zannoni [6], is based on the formation of molecular clusters with internal orientational ordering that persists in the isotropic phase as well. The ordering within the clusters is dictated by the short-range anisotropic interactions whereas the macroscopic nematic order is dominated by the long-range anisotropic interactions among the clusters. The shape of the clusters need not be as anisometric as that of the individual molecules and therefore the isotropic averaging of just the long-range interactions over the inter-cluster positions would not be unrealistic. A similar justification, invoking *"sterically determined"* short range order, was proposed by Maier and Saupe [3] in the original development of their theory, where they suggested that bundles of molecules, rather than individual molecules, could be considered as the interacting units for the statistical treatment. For the case of azoxyanisole, they estimated that such bundles consist of approximately four molecules arranged parallel to one another.

There were early reports on experimental observations of possible ferroelectric effects in common nematics [10] suggesting the presence of ordered molecular aggregates in both, the nematic and the isotropic phases. Direct X-ray manifestations of the presence of clusters in certain nematics, termed by De Vries as cybotactic, were reported and confirmed in the early 70's [11] and are now commonly observed in many types of nematogens, either as pretransitional manifestations of a lower temperature smectic phase or as stable structural features. Analogous structures where known long before in normal liquids [12], for which the term "cybotactic" was originally introduced. Lately, it is becoming increasing clear [13-20] that the presence of local ordered structures is the key to understanding many of the properties of the thermotropic nematic phase, particularly the manifestations of biaxial and polar ordering. As succinctly put by Samulski [20] "all nematics are cybotactic". It seems therefore appropriate to attempt to formulate a "simple theory of all nematics" by extending the M-S theory to explicitly allow for the presence of internally ordered molecular clusters. This is the purpose of the present paper.

After a brief review of the M-S theory, the formulation of the free energy for the extended theory is outlined in the next section. For simplicity, the formulation is restricted to uniaxial molecules, and therefore to uniaxial nematics. However, the extended theory would be particularly useful for the description of biaxial nematics, since it extends Freiser's theory of biaxial nematics [21] to the experimentally more relevant regime of cybotactic nematics consisting of biaxial clusters [13-20]. A Landau-de Gennes description of the latter type of nematics has been reported [13]. Results of calculations on the molecular ordering and on the thermodynamics of the phase transitions involved are presented in section 3 and are discussed relative to the results of the M-S theory. Section 4 contains the conclusions and the perspectives of the extended theory.

**2. Internal and macroscopic nematic order of molecular clusters in the mean-field approximation.**

*2.1 Review of the Maier-Saupe theory.*

The mean-field character of the M-S theory is due to the neglect of correlations between the orientations of neighboring molecules in the nematic liquid [22]. Thus, a molecule is assumed to orient under the action of a mean-field, independently of the orientations of its neighbors. In turn, the mean field originates from the collective alignment of all the molecules surrounding that molecule and therefore reflects the extent of molecular ordering in the phase. In statistical mechanics terminology, the mean-field approximation (MFA) entails the replacement of the $N$-molecule joint probability distribution $p^{(N)}$ by a product of $N$ effective probability distributions $f(\omega_i)$ of the orientations of each of the $N$ molecules. Schematically,

$$p^{(N)}(\mathbf{r}_1,\omega_1;\mathbf{r}_2,\omega_2;....\mathbf{r}_N,\omega_N) \xrightarrow{MFA} V^{-N} f(\omega_1) \cdot f(\omega_2) \cdot ... \cdot f(\omega_N), \qquad (1)$$

where $N$ is the number of molecules in the nematic liquid, $\omega_i$ is the orientation of the $i^{th}$ molecule ($i=1,2…N$) relative to the nematic director **n**, and $\mathbf{r}_i$ is the position vector of that molecule within the volume $V$ of the sample.

For molecules interacting in pairs, the effective probability distribution can be related self-consistently to the intermolecular pair potential $u(\mathbf{r}_{i,j},\omega_i,\omega_j)$ by minimization, at constant density, of the free energy functional

$$\frac{F}{N} = \frac{1}{2}(N-1)\int f(\omega_1)f(\omega_2)\bar{u}(\omega_1,\omega_2)d\omega_1 d\omega_2 + k_B T \int f(\omega)\ln f(\omega)d\omega. \qquad (2)$$

Here, the positionally averaged anisotropic potential $\bar{u}(\omega_1,\omega_2)$ is given by the exact (i.e. not restricted to the MFA) expression

$$\bar{u}(\omega_1,\omega_2) = \frac{1}{V}\int g(\mathbf{r}_{1,2},\omega_1,\omega_2) u(\mathbf{r}_{1,2},\omega_1,\omega_2) d\mathbf{r}_{1,2}, \qquad (3)$$

where $g(\mathbf{r}_{1,2},\omega_1,\omega_2)$ is the pair correlations function between molecules 1 and 2. In the M-S theory the molecules are assumed, for simplicity, to be perfectly symmetric about their long axis (uniaxial molecules) and a crucial approximation is made by putting $\bar{u}$ in the form

$$\bar{u}(\vartheta_1,\vartheta_2) \approx -u_0 \frac{a^3}{V} P_2(\cos\vartheta_{1,2}) \qquad (4)$$

in which $u_0$ and $a$ represent, respectively, effective strength and range parameters of the anisotropic part of molecular interaction. The angles $\vartheta_1,\vartheta_2$ describe the orientations of the long axes of molecules 1 and 2 relative to the director **n,** $\vartheta_{1,2}$ denotes the angle of those axes relative to each other and $P_2(\cos\vartheta_{1,2}) \equiv (3\cos^2\vartheta_{1,2}-1)/2$ is the second Legendre polynomial. Using the approximate anisotropic potential of Equation (4), the functional minimization of the free energy in Equation (2) with respect to the orientational distribution $f(\vartheta)$ leads to the well known expression

$$f(\vartheta) = \frac{1}{\zeta} \exp(bSP_2(\cos\vartheta)) \quad, \tag{5}$$

in which

$$b = \frac{u_0}{k_B T} \frac{(N-1)a^3}{V} \tag{5'}$$

is a dimensionless inverse temperature coefficient that depends on the strength and range of the anisotropic part of the molecular interactions and on the density. The normalization factor is $\zeta \equiv \int_{-1}^{1} \exp(bSP_2(\cos\vartheta))d(\cos\vartheta)$ and $S$ is the so called Saupe order parameter of the nematic phase, which is evaluated in terms of $b$ by solving the self-consistency equation

$$S \equiv \int_{-1}^{1} f(\vartheta) P_2(\cos\vartheta) d(\cos\vartheta). \tag{6}$$

When this equation is satisfied, the free energy expression in Equation (2) leads to the equation

$$\frac{F - F_0}{Nk_B T} = \frac{1}{2} bS^2 - \ln(\zeta/2) \quad, \tag{7}$$

in which $F_0$ denotes the fully isotropic part of the free energy. Equation (6) has the solution $S = 0$, corresponding to the isotropic phase, and, for $b$ larger than a critical value $b_c \simeq 4.48$, it also has solutions with $S \neq 0$, corresponding to the nematic phase. According to the free energy expression in Equation (7), the nematic phase solutions of Equation (6) become thermodynamically stable for $b$ above the value $b_{N-I} \simeq 4.54$, which therefore marks the nematic to isotropic (*N-I*) transition. At the transition, the order parameter changes discontinuously from $S = 0$ to the value $S_{N-I} = 0.43$, from which it increases continuously with further increasing $b$, as sown in the well known diagram of Figure 1. According to Equation (7), the entropy change $\Delta s$ per molecule at the transition is fixed to $(\Delta s)_{N-I} / Nk_B = -b_{N-I} S_{N-I}^2 / 2 \simeq -0.42$.

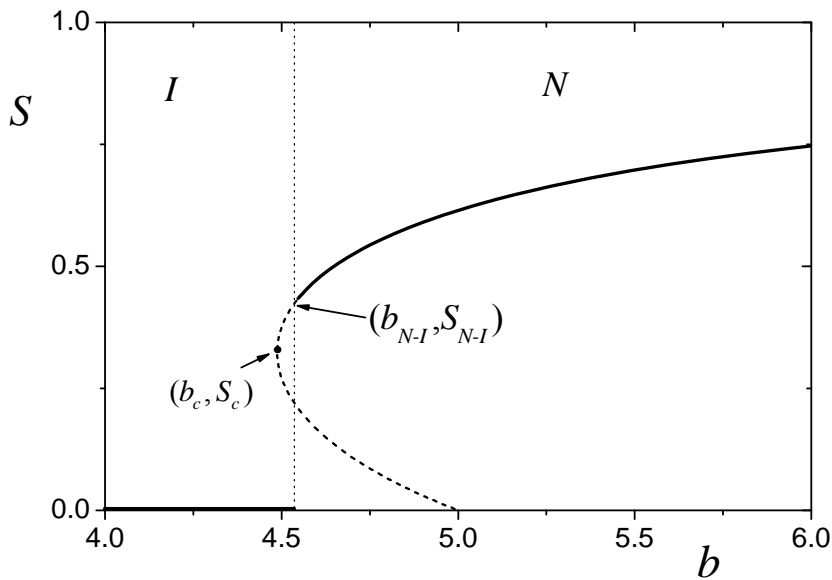

**Figure 1.** Order parameter vs *b* (inverse temperature) according to the M-S theory. The solid line corresponds to the thermodynamically stable states and the dotted one to ordered solutions of the self consistency equation which lack thermodynamic stability. The phase transition point $(b_{N-I}, S_{N-I})$ and the onset $(b_c, S_c)$ of ordered solutions upon cooling are also indicated.

*2.2 Internal and macroscopic nematic ordering of molecular clusters.*

After this brief review of the M-S theory, we outline the formulation of the free energy for a system of uniaxial molecules organized into clusters. As a conceptual starting point we may consider the replacement of each individual molecule in the M-S formulation by a cluster of molecules. In this picture, the direction of the long axis of the M-S molecule is replaced by the direction of preferential alignment of the molecules forming the cluster. Equivalently, one may consider the partitioning of the entire sample of the *N* molecules into a number *R* of clusters, with the molecules of each cluster exhibiting preferential alignment along some direction (see Figure 2). In any case, the ordering within the clusters is not assumed to be perfect. Accordingly each cluster is characterized not only by the number of molecules forming it and by the direction of preferential alignment of those molecules, but also by the degree of alignment of the molecules and by the energy and entropy associated with that degree of alignment. Thus, part of the free energy of the entire sample is due to the internal interactions and to the ordering within the individual clusters. Another part comes from the collective energy and entropy of the clusters, which interact with one another and also move and reorient relative to one another. A third part of the free energy is purely entropic and is due to the liquid nature of the system which allows the exchange of molecules among clusters.

To put these parts together, lets label each cluster by the index $r$ (=1,2,...R) and denote by $n_r$ the number of molecules forming the $r^{th}$ cluster. The sum of the cluster populations gives the total number $N$ of molecules in the sample, i.e. $\sum_{r=1}^{R} n_r = N$. Using the unit vector $\mathbf{n}^{(r)}$ to denote the direction of preferential alignment (local director) of the molecules within the $r^{th}$ cluster, the orientation of that cluster relative to the macroscopic director $\mathbf{n}$ of the nematic phase can be specified by the angle $\Theta_r$ (see Figure 2), where $\cos\Theta_r = \mathbf{n} \cdot \mathbf{n}^{(r)}$. The molecules within the $r^{th}$ cluster are labeled by the index $i_r = 1,2...n_r$, and the directions of their long axes relative to the local director $\mathbf{n}^{(r)}$ are specified by the angles $\theta_{i_r}$. The local director is defined as the principal $z$ axis of the tensor $S'^{(r)}_{ab} \equiv \frac{1}{n_r} \sum_{i_r=1}^{n_r} \left(3(\mathbf{l}_{i_r} \cdot \mathbf{a})(\mathbf{l}_{i_r} \cdot \mathbf{b}) - (\mathbf{a} \cdot \mathbf{b})\right)/2$, where $\mathbf{l}_{i_r}$ denotes the unit vector associated with the long axis of the $i_r$ molecule and $\mathbf{a}, \mathbf{b}$ stand for unit vectors in the directions of the cluster-fixed axes $x^{(r)}, y^{(r)}, z^{(r)}$. Accordingly, the averaging of the molecular orientations within a cluster is subject to the constraints

$\sum_{i_r=1}^{n_r} \cos\varphi_{i_r} \sin 2\theta_{i_r} = \sum_{i_r=1}^{n_r} \sin\varphi_{i_r} \sin 2\theta_{i_r} = \sum_{i_r=1}^{n_r} \sin 2\varphi_{i_r} \sin^2\theta_{i_r} = 0$ for the angles $\theta_{i_r}, \varphi_{i_r}$ of the molecular vectors $\mathbf{l}_{i_r}$ in the cluster-fixed frame $x^{(r)}, y^{(r)}, z^{(r)}(\|\mathbf{n}^{(r)})$. Similarly, the macroscopic director $\mathbf{n}$ is defined as the $Z$ principal axis of the tensor $S''_{AB} \equiv \frac{1}{N} \sum_{r=1}^{R} n_r \left(3(\mathbf{n}^{(r)} \cdot \mathbf{A})(\mathbf{n}^{(r)} \cdot \mathbf{B}) - (\mathbf{A} \cdot \mathbf{B})\right)/2$, with $\mathbf{A}, \mathbf{B}$ denoting the directions of the macroscopic axes $X, Y, Z$. Hence, the averaging of the cluster orientations is subject to analogous constraints for the angles $\Theta_r, \Phi_r$ that describe the directions of the cluster directors $\mathbf{n}^{(r)}$ in the macroscopic frame.

To simplify the formulation of the free energy, we assume that all the clusters are of the same size, i.e. that $n_r \simeq n = N/R$ for all $r=1,2...R$. At uniform molecular density $\rho = N/V$, the assumption of equal populations implies that the clusters are of equal volume $\upsilon \equiv n/\rho = V/R$ and furthermore that they have the same degree of internal ordering (described by the single order parameter $S'^{(r)}_{zz} = S'$), albeit in different directions $\mathbf{n}^{(r)}$, and also the same degree of ordering with respect to the macroscopic director $\mathbf{n}$ (described by the order parameter $S'' = S''_{ZZ}$).

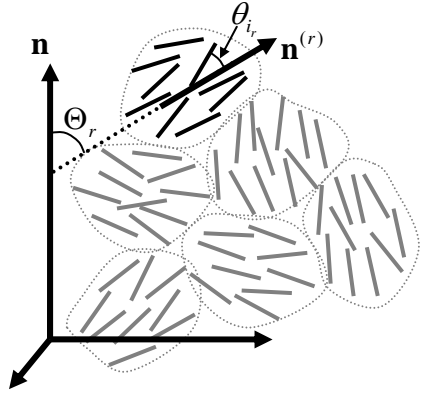

**Figure 2.** Schematic representation of the internal molecular order in the clusters and of their relative orientations. Line segments represent the long axes of the molecules. $\Theta_r$ is the polar angle between the local director of the $r^{th}$ cluster, $\mathbf{n}^{(r)}$, and the macroscopic nematic director $\mathbf{n}$. $\theta_{i_r}$ is the angle between the long axis of a molecule and the local director of the cluster it belongs to.

In order to apply the MFA of Equation (1) to the ensemble of molecular clusters illustrated in Figure 2, we introduce two types of orientational distribution functions. For a molecule $i$ belonging to the $r^{th}$ cluster we use $f'(\theta_{i_r})$ to describe the distribution of its molecular axis relative to the cluster director $\mathbf{n}^{(r)}$. The other type of distribution function is denoted by $f''(\Theta_r)$ and describes the distribution of the cluster director $\mathbf{n}^{(r)}$ relative to the macroscopic director $\mathbf{n}$. In accord with the assumption of identical cluster sizes, what changes in the orientational distributions $f'$ and $f''$ on going from one cluster to the other is just the orientation of the cluster director $\mathbf{n}^{(r)}$. Then, in analogy with Eq(1), the orientational entropy in the MFA can be expressed as the sum of two contributions: One is associated with the reorientations of the molecules within the clusters and can be determined from the respective orientational distribution as $-k_B R(n-1)\int f'(\theta)\ln(f'(\theta))d(\cos\theta)$. The other contribution is associated with the reorientations of the cluster directors $\mathbf{n}^{(r)}$ and is given by $-k_B (R-1)\int f''(\Theta)\ln(f''(\Theta))d(\cos\Theta)$. Note that although there are R clusters in the sample, the reorientations of their respective directors are not completely independent, due to the constraints involved in the definition of the macroscopic director. Hence the factor $(R-1)$ in the orientational entropy of the $R$ clusters. Similarly, the factor $(n-1)$ in the orientational entropy of the $n$ molecules within any of the $R$ clusters is to account for the constraints imposed by the identification of the respective cluster director.

The free energy contribution of the anisotropic interactions is evaluated in analogy with Eqs (3) and (4) and can be separated into terms associated with interactions among molecules belonging to the same cluster and interactions among molecules belonging to different clusters. The former are given by

$$\bar{u}'(\theta_{1,2}) = \frac{1}{\upsilon^2}\int_\upsilon d\mathbf{r}_2 \int_\upsilon d\mathbf{r}_2 g(\mathbf{r}_{1,2},\theta_1,\theta_2)u(\mathbf{r}_{1,2},\theta_1,\theta_2) \simeq -u_0 \frac{a^3}{\upsilon}v' P_2(\cos\theta_{1,2}) \tag{8}$$

and the latter by

$$\bar{u}''(\vartheta_{1_r,1_{r'}}) = \frac{1}{V\upsilon^2}\int_V \left(\int_{\upsilon_{(r')}}\int_{\upsilon_{(r)}} g(\mathbf{r}_{1_r,1_{r'}},\vartheta_{1_r},\vartheta_{1_{r'}})u(\mathbf{r}_{1_r,1_{r'}},\vartheta_{1_r},\vartheta_{1_{r'}})d\mathbf{r}_{1_r}d\mathbf{r}_{1_{r'}}\right)d\mathbf{R}_{r,r'}$$
$$\simeq -u_0 \frac{a^3}{V}v'' P_2(\cos\vartheta_{1_r,1_{r'}}) \tag{9}$$

where $\mathbf{R}_{r,r'}$ denotes the position of cluster $r'$ relative to cluster $r$ and the integrations over the molecular positions $\mathbf{r}_{1_r}$ and $\mathbf{r}_{1_{r'}}$ extend respectively over the volumes of the clusters $r$ and $r'$. The dimensionless factors $v'$ and $v''$ are functions of the cluster volume. Therefore, at uniform molecular density they are functions of the cluster population $n$. Obviously the in-cluster integration factor $v'$ vanishes for $n \leq 1$ and reaches the asymptotic value $v' \to 1$ for macroscopically large clusters ($n \to N$). The $n$-dependence of $v'$ for intermediate cluster sizes depends on the details of the molecular interactions and, to some extent, on the geometry of the clusters. The shorter the effective range of the intermolecular interactions the more rapidly $v'$ approaches the saturation value 1. An explicit form of this dependence is considered in section **3**. Turning now to $v''$ of Equation (9), we note that at uniform molecular density the spatial average of the interaction between a pair of molecules should be fixed at the value given in eq (4) independently of which clusters the molecules belong to. Therefore, combining Equation (4) with (8) and (9), we obtain the general relation $v' + v'' = 1$.

Altogether then, for uniform molecular density $N/V$ and uniform number $n$ of molecules per cluster, the relevant free energy difference is

$$\frac{F-F_0}{Nk_BT} = -\frac{1}{2}bv'\int f'(\theta_1)f'(\theta_2)P_2(\theta_{1,2})d(\cos\theta_1)d(\cos\theta_2) + \frac{n-1}{n}\int f'(\theta)\ln(2f'(\theta))d(\cos\theta)$$
$$-\frac{1}{2}b(1-v')\int f''(\Theta_r)f''(\Theta_{r'})f'(\theta_{1_r})f'(\theta_{1_{r'}})P_2(\vartheta_{1_r,1_{r'}})d(\cos\theta_{1_r})d(\cos\theta_{1_{r'}})d(\cos\Theta_r)d(\cos\Theta_{r'})$$
$$+\frac{R-1}{N}\int f''(\Theta)\ln(2f''(\Theta))d(\cos\Theta)$$

(10)

The functional minimization of this free energy with respect to $f'$ and $f''$, leads to the following expressions for these distribution functions,

$$f'(\theta) = \frac{1}{\zeta'}e^{b\frac{n}{n-1}(v'+(1-v')S''^2)S'P_2(\cos\theta)},$$
$$f''(\Theta) = \frac{1}{\zeta''}e^{b\frac{N}{R-1}(1-v')S'^2S''P_2(\cos\Theta)}.$$

(11)

The order parameters $S'$ and $S''$ measuring, respectively, the degree of molecular ordering within the clusters and the degree of ordering of the clusters in the macroscopic sample, are obtained from the self consistency conditions

$$S' = \int_{-1}^{1}f'(\theta)P_2(\cos\theta)d(\cos\theta),$$
$$S'' = \int_{-1}^{1}f''(\Theta)P_2(\cos\Theta)d(\cos\Theta).$$

(12)

When these conditions are satisfied, the free energy of Eq(10) can be expressed as

$$\frac{F-F_0}{Nk_BT} = \frac{1}{2}b(v'+3(1-v')S''^2)S'^2 - \frac{n-1}{n}\ln(\zeta'/2) - \left(\frac{1}{n}-\frac{1}{N}\right)\ln(\zeta''/2) \quad (13)$$

A third condition, from which the cluster size parameter $n$ can be specified in terms of the temperature variable $b$, is obtained by minimizing the free energy with respect to $n$, subject to the constraint $nR=N=$constant $\gg 1$. This yields the equation

$$bS'^2n\left(\frac{1}{2}(1-S''^2)n\frac{\partial v'}{\partial n} + \frac{R}{R-1}(1-v')S''^2 - \frac{v'+(1-v')S''^2}{n-1}\right) = \ln(\zeta''/\zeta') \quad (14)$$

It is apparent from Equations (10) to (14) that, in the limit $n \to N$ (therefore $R \to 1$, i.e. a sample consisting of a single cluster that contains all the molecules and $v' \to 1$) the distribution $f''$ is irrelevant since the director $\mathbf{n}^{(1)}$ of the single cluster coincides with the macroscopic director $\mathbf{n}$, implying $S'' = 1$ and the distribution $f'$ becomes identical to the M-S distribution $f$ of Equation (5). Similarly in the limit $n \to 1$ (therefore $R \to N$ i.e. the clusters are identified with single molecules and $v' \to 0$) $f'$ becomes irrelevant since the director $\mathbf{n}^{(r)}$ of the single cluster coincides with molecular axis $\mathbf{l}$, and the distribution $f''$ becomes identical to $f$ of the M-S theory. In both limiting cases the free energy expressions in eqs (7) and (13) coincide.

**3. Results and discussion.**

The free energy minimization conditions in Equation (12) and (14) accept three kinds of solutions corresponding to:

**(i)** The "molecular" isotropic phase (*I*), in which $S' = S'' = 0$. The subdivision into clusters in this phase is meaningless as, in the absence of internal ordering, a cluster director $\mathbf{n}^{(r)}$ cannot be defined. Therefore, cluster size is irrelevant for this phase.

**(ii)** The "cybotactic" isotropic phase (*I′*), in which $S' \neq 0, S'' = 0$, consisting of internally ordered clusters whose orientations are distributed, yielding a macroscopically isotropic fluid. According to Equations (12-13), for sufficiently large $n$ so that $(n-1)/n \simeq 1$, the self-consistency conditions and the free energy for the transition from the *I* to the *I′* phase differ from those of the *N-I* transition in the M-S theory only in that $b$ is replaced by $b' = bv'$. Accordingly, the *I′* phase is stabilized relative to the I phase for $b'$ exceeding the universal value $b'_{I'-I} \simeq 4.54$. This corresponds to a transition temperature $T_{I'-I}$ which is higher than the temperature $T_{N-I}^{(M-S)}$, predicted for the $N-I$ transition in the M-S theory, by a factor $T_{I'-I}/T_{N-I}^{(M-S)} = (v')^{-1} > 1$. Also, the solutions corresponding to the *I′* phase disappear for $b' < b_c \simeq 4.48$. Aside from that, the in-cluster order parameter $S'$ changes at the $I'-I$ transition discontinuously from $S' = 0$ to the universal value $S'_{I'-I} = 0.43$ and follows a $b'$-dependence $S'(b')$ which is the exact analogue of the M-S $S(b)$ shown in Figure 1.

**(iii)** The nematic phase (*N*), in which $S', S'' \neq 0$. Depending on the functional form of $v'$, the *N* phase can either be obtained directly from the *I* phase on increasing *b* or can evolve from the $I'$ phase. In the latter case the internally ordered clusters merge, on lowering the temperature, to form a single cluster of the same uniaxial nematic order and of macroscopic size ($n \to N$). The stabilisation of the *N* phase relative to $I'$, when applicable, is obtained above a value of *b* which, unlike the M-S theory, is not universal and depends on $v'$. According to its definition in Equation (6), the molecular order parameter *S* for the macroscopic nematic phase is obtained by averaging the orientations of a molecule relative to the macroscopic nematic director **n**. Its value is therefore equal to the product of the order parameters $S'$ (order of the molecule relative to the cluster director $\mathbf{n}^{(r)}$) and $S''$ (order of $\mathbf{n}^{(r)}$ relative to **n**), i.e. $S = S'S''$.

To describe the possible phase transitions in more detail we consider a specific form of the function $v'(n)$. Rather than choosing a particular pair potential for the molecular interaction and evaluating $v'(n)$ using Equation (8), we will choose directly the functional form of $v'(n)$ subject only to the general requirements that it should be a continuous function of $n \geq 1$ increasing monotonously from 0 (at $n = 1$) to 1 (as $n \to \infty$). For the present illustrative purposes, a simple generic form that meets these requirements is given by the exponential dependence $v'_E(n) = e^{-[\alpha/(n-1)]^\gamma}$, with the constants $\alpha, \gamma > 0$. As shown below, this form can lead to different phase sequences depending on the values of the parameters $\alpha$ and $\gamma$, which illustrates adequately the influence of $v'(n)$, and thereby of the molecular interactions, on the formation of ordered microstructures.

The results for case with $\gamma = 1$ are summarized in Figures 3 and 4. For $\alpha > 1$ the description is equivalent to the M-S theory: Only a direct transition from the *I* to the *N* phase is obtained on increasing *b*. The transition temperature as well order parameter and entropy values coincide with the universal values of the M-S theory. For $\alpha < 1$, the $I'$ phase appears in the diagram (Figure 3a) at intermediate temperatures between *I* and *N*. The transition values $b_{I'-I}$ and $b_{N-I'}$ respectively increase and decrease continuously with decreasing $\alpha$, staring out from the common universal value $b_{N-I}$ at $\alpha = 1$.

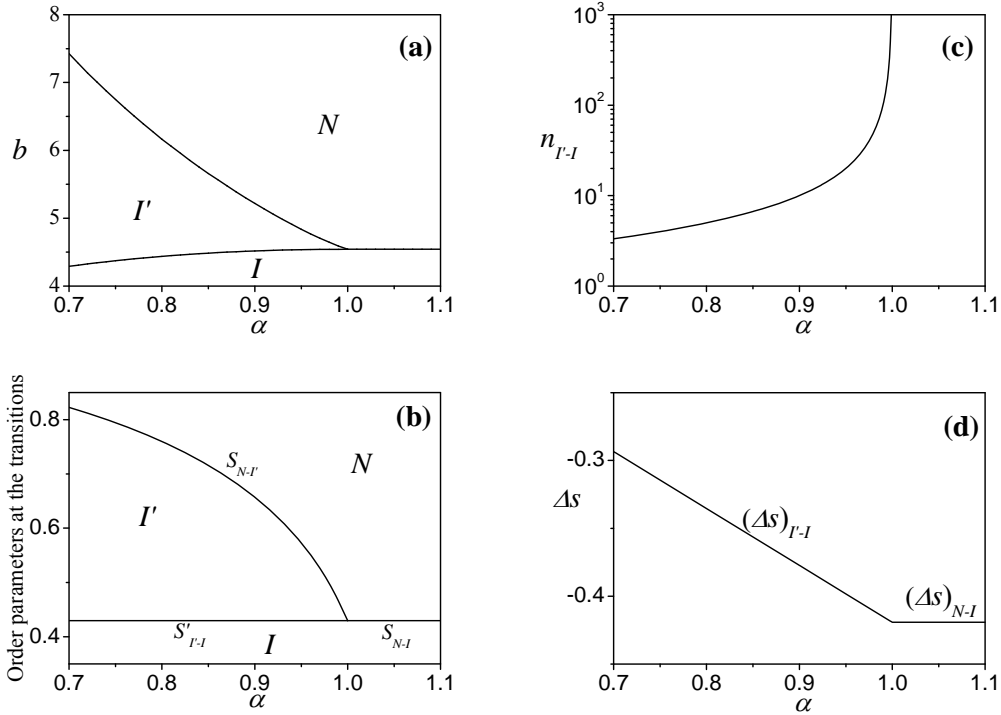

**Figure 3.** Calculated phase-transition diagrams using the functional form $v'_E(n) = e^{-\alpha/(n-1)}$ for the in-cluster integration factor $v'$ of Equation (8) with variable $\alpha$. (a) Regions of phase stability and values of the inverse effective temperature $b$ at the transitions among the $I$, $I'$ and $N$ phases. (b) Order parameter values at the phase transitions. (c) Cluster population at the $I'$-$I$ transition, in logarithmic scale. (d) Changes of the orientational entropy per particle (in units of $k_B$) at the transitions from the $I$ phase.

The values of the cluster order parameter $S'$ at the transition from the $I$, either to the $N$ or to the $I'$ phases, remain fixed to the universal value $S'_{I'-I} = S'_{N-I} = 0.43$ independently of $\alpha$, whilst the value of $S'$ at the $N - I'$ increases continuously with decreasing $\alpha < 1$ (Figure 3b). This is generally in accord with the relatively high order observed in cybotactic nematics [17, 23, 24] and reflects the fact that the macroscopic ordering results from the alignment of already ordered bundles of molecules, rather than individual molecules. In contrast, the fixed low value of $S'_{I'-I}$ reflects the fact that the $I' - I$ transition results from the ordering of individual molecules.

The cluster population $n$ undergoes a discontinuous jump at the $I - I'$ transition from $n = 1$ in the $I$ to $n = n_{I'-I}$ in the $I'$ phase and grows from there continuously with decreasing temperature (see Figure 4). The transition values $n_{I'-I}$ start out from diverging values at $\alpha = 1$ and decrease with decreasing $\alpha$ (Figure 3c). The entropy change $(\Delta s)_{I'-I}$ at the $I-I'$ transition is related to the M-S universal value of the entropy change $(\Delta s)_{N-I}^{(M-S)}$ according to relation $(\Delta s)_{I'-I} = \alpha (\Delta s)_{N-I}^{(M-S)}$ and therefore decreases continuously with decreasing $\alpha$ (Figure 3d). The entropy change for the $N-I'$ is found to vanish.

The temperature dependence of the molecular order parameters $S'$ and $S$ and of the cluster population $n$ are shown in Figure 4 for the particular value of the parameter $\alpha = 0.95$. It is clear from these diagrams that the stabilization of the $N$ phase occurs as the cluster population grows divergently large and that there is no discontinuous change of the order parameter.

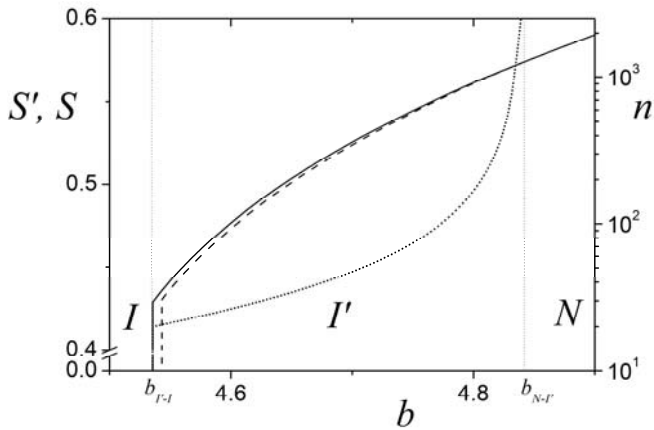

**Figure 4.** Calculated $b$-dependence of order parameters and cluster populations using the same functional form as in Figure 3 for $v'$ with the parameter $\alpha$ fixed at $\alpha = 0.95$. i) Order parameters $S'$ (solid line for $b < b_{N-I'} \simeq 4.84$) and $S$ (solid line for $b \geq b_{N-I'}$), and ii) cluster size $n$ (dotted line) in logarithmic scale. At $b_{N-I'}$ $S = S' \simeq 0.573$. The dashed line corresponds to the order parameter $S$ of the nematic phase in its metastable temperature range.

As $n$ increases with $b$ approaching $b_{N-I'}$, the free energy of the nematic phase gets only slightly higher than that of the $I'$ phase and therefore stabilization of $N$ in that case can be achieved by applying a weak aligning field to the $I'$ phase. Thus, in the large-$n$ regime, the $I'$ phase shows the behavior of conventional thermotropic nematics which form ordered domains of macroscopic size that can be field-aligned into a state of uniform director.

Lastly, it should be noted that not only the quantitative features but also some of the qualitative trends shown in Figures 3 and 4 for $\gamma = 1$ do change on varying the value of the $\gamma$ parameter. However, such variations will not be discussed here.

## 4. Conclusions and perspectives

Our formulation of the molecular theory of cybotactic nematics has followed closely the Maier-Saupe theory in that it is based on the mean field approximation and uses the same functional form for the orientation dependence of the effective intermolecular potential. The essential difference is that the possibility of the formation of molecular clusters with internal orientational order is explicitly allowed for. In addition to its very profound implications on the nature of the nematic phase, this extension of the theory broadens significantly the range of predictions and provides a systematic and physically clear way of relaxing some of the inherent restrictions of the Maier-Saupe theory regarding the universal values for the order parameter and the entropy change at the *N-I* transition. Specifically:

(i) The theory predicts two stable macroscopically isotropic phases, one with complete molecular disorder ($I$) and the other ($I'$) consisting of internally ordered molecular clusters. The two phases are connected by a first order phase transition. The theory also allows for phase transitions from either of these phases to the macroscopically ordered nematic phase *N*.

(ii) The stability of the above phases at a given temperature is dictated by the detailed intermolecular position dependence the anisotropic interactions.

(iii) The cluster size in the $I'$ phase increases continuously with decreasing temperature; on approaching the transition temperature to the nematic phase it diverges to macroscopic values.

(iv) Depending on the intermolecular potential, the values of the transition temperatures, order parameters and entropy changes may show considerable deviations from the universal values predicted by the Maier-Saupe theory.

(v) The Maier-Saupe theory appears as a particular case of the extended theory, obtained in the limits of single-molecule clusters or of macroscopic samples consisting of a single cluster.

The extended theory retains the simplicity of the Maier-Saupe theory. It also retains its main weaknesses, notably the neglect of molecular correlations and the absence of possible transitions to ordered phases other than the nematic. Additionally, the present simplified formulation of the theory neglects the possible dispersion in the cluster sizes. More importantly, the internal ordering of the clusters is restricted to have the same symmetry with the ordered phase, i.e. uniaxial nematic, thus excluding the experimentally interesting cases where the clusters have smectic internal ordering and give rise to macroscopically nematic phases. However, a more general formulation is possible [25] wherein these additional restrictions can be relaxed. This opens up new perspectives for the understanding of subtle features of the nematic phase stemming from the possibility of molecular self-organization in a hierarchy of ordered microstructures. Such microstructures, not necessarily of the nematic type, can yield macroscopically nematic media with significantly different physical properties and symmetries [14, 18, 20, 26] from the "molecular" nematics described in the pioneering work of Maier and Saupe and its subsequent generalization to biaxial nematics [21].

**Acknowledgements**
Part of this research has been funded through the EU 7$^{th}$ Framework Programme (FP7/2007-2013) under the project BIND (Biaxial Nematic Devices, grant agreement #216025).

**References**

1. Maier, W.; Saupe, A. *Z. Naturforsch.* **1958**, *13a*, 564.
2. Maier, W.; Saupe, A. *Z. Naturforsch.* **1959**, *14a*, 882.
3. Maier, W.; Saupe, A. *Z. Naturforsch.* **1960**, *15a*, 287.
4. Ypma, J.G.Y.; Vertogen, G.; Koster, H.T. *Mol. Cryst. Liq. Cryst.* **1976**, *37*, 57–69.
5. Humphries, R.L.; James, P.G.; Luckhurst, G.R. *J. Chem. Soc. Faraday Trans. II* **1972**, *68*, 1031–1039.
6. Luckhurst, G.R.; Zannoni, C. *Nature* **1977**, *267*, 412–414.
7. De Jeu, W.H. *Mol. Cryst. Liq. Cryst.* **1997**, *292*, 13–24.
8. Gelbart, W.M.; Gelbart, A. *Mol. Phys.* **1977**, *33*, 1387–1399.

9.  Gelbart, W.M.; Barboy, B. *Acc. Chem. Res.* **1980**, *13*, 290–296.
10. Williams, R.; Heilmeier, G. *J. Chem. Phys.* **1966**, *44*, 638.
11. De Vries, A. *Mol. Cryst. Liq. Cryst.* **1970**, *10*, 31; *ibid* 219; *ibid* **1973** *20*, 119.
12. Stewart, G.W.; Morrow, R.M. *Phys. Rev.* **1927**, *30*, 232–244.
13. Vanakaras, A.G.; Photinos, D.J. *J. Chem. Phys.* **2008**, *128*, 154512.
14. Peroukidis, S.D.; Karahaliou, P.K.; Vanakaras, A.G.; Photinos, D.J. *Liq. Cryst.* **2009**, *36*, 727–737.
15. Francescangeli, O.; Stanic, V.; Torgova, S.I.; Strigazzi, A.; Scaramuzza, N.; Ferrero, C.; Dolbnya, I.P.; Weiss, T.M.; Berardi, R.; Muccioli, L.; Orlandi, S.; Zannoni, C. *Adv. Funct. Mater.* **2009**, *19*, 2592–2600.
16. Görtz, V.; Southern, C.; Roberts, N.W.; Gleeson, H.F.; Goodby, J.W. *Soft Matter* **2009**, *5*, 463–471.
17. Keith, C.; Lehmann, A.; Baumeister, U.; Prehm, M.; Tschierske, C. *Soft Matter*, DOI: 10.1039/b923262a.
18. Tschierske, C.; Photinos, D. J, *J. Mat. Chem.* **2010** (in press).
19. Vaupotič, N.; Szydlowska, J.; Salamonczyk, M.; Kovarova, A.; Svoboda, J.; Osipov, M.; Pociecha, D.; Gorecka, E. *Phys. Rev. E* **2009**, *80*, 030701.
20. Samulski, E.T. *Liq. Cryst.* (this issue).
21. Freiser, M. J. *Phys. Rev. Lett.* **1970**, *24*, 1041–1043.
22. An excellent pedagogical presentation of the M-S theory can be found in *The Molecular Physics of Liquid Crystals*, Luckhurst, G. R., Gray, G.W., Eds; Academic Press, 1979; Chapter 4. See also de Gennes, P.G.; Prost, J. *The Physics of Liquid Crystals*, Oxford University Press, 1993.
23. Karahaliou, P. K.; Kouwer, P. H. J.; Meyer, T.; Mehl, G. H.; Photinos, D. J. *J. Phys. Chem. B.* **2008**, *112*, 6550–6556.
24. Karahaliou, P. K.; Kouwer, P. H. J.; Meyer, T.; Mehl, G. H.; Photinos, D. J. *Soft Matter* **2007**, *3*, 857–865.
25. Droulias, S.; Vanakaras, A.G.; Photinos, D.J., to be published.
26. Karahaliou, P. K.; Vanakaras, A. G.; Photinos, D. J. *J. Chem. Phys.* **2009**, *131*, 124516.

**FIGURE CAPTIONS**

**Figure 1.** Order parameter vs $b$ (inverse temperature) according to the M-S theory. The solid line corresponds to the thermodynamically stable states and the dotted one to ordered solutions of the self consistency equation which lack thermodynamic stability. The phase transition point $(b_{N-I}, S_{N-I})$ and the onset $(b_c, S_c)$ of ordered solutions upon cooling are also indicated.

**Figure 2.** Schematic representation of the internal molecular order in the clusters and of their relative orientations. Line segments represent the long axes of the molecules. $\Theta_r$ is the polar angle between the local director of the $r^{th}$ cluster, $\mathbf{n}^{(r)}$, and the macroscopic nematic director $\mathbf{n}$. $\theta_{i_r}$ is the angle between the long axis of a molecule and the local director of the cluster it belongs to.

**Figure 3.** Calculated phase-transition diagrams using the functional form $v'_E(n) = e^{-\alpha/(n-1)}$ for the in-cluster integration factor $v'$ of Equation (8) with variable $\alpha$. (a) Regions of phase stability and values of the inverse effective temperature $b$ at the transitions among the *I, I'* and *N* phases. (b) Order parameter values at the phase transitions. (c) Cluster population at the *I'-I* transition, in logarithmic scale. (d) Changes of the orientational entropy per particle (in units of $k_B$) at the transitions from the *I* phase.

**Figure 4.** Calculated $b$-dependence of order parameters and cluster populations using the same functional form as in Figure 3 for $v'$ with the parameter $\alpha$ fixed at $\alpha = 0.95$. i) Order parameters $S'$ (solid line for $b < b_{N-I'} \simeq 4.84$) and $S$ (solid line for $b \geq b_{N-I'}$), and ii) cluster size $n$ (dotted line) in logarithmic scale. At $b_{N-I'}$ $S = S' \simeq 0.573$. The dashed line corresponds to the order parameter $S$ of the nematic phase in its metastable temperature range.